\documentclass[aps,prd,eqsecnum,showpacs]{revtex4}
\usepackage{color,graphicx}
\usepackage{subfigure}
\usepackage{amsfonts}
\usepackage{amssymb}
\begin{document}


\title{
Asymptotically Friedmann self-similar scalar field solutions with 
potential
}

\author{
$^{1}$Masanori Kyo, 
$^{1}$Tomohiro Harada\footnote{harada@rikkyo.ac.jp}
and 
$^{2,3}$Hideki Maeda\footnote{hideki@cecs.cl}}
\affiliation{
$^{1}$Department of Physics, Rikkyo University, Toshima, Tokyo 171-8501,
Japan\\$^{2}$Centro de Estudios Cient\'{\i}ficos (CECS), Arturo Prat 514, 
Valdivia, Chile\\
$^{3}$Department of Physics, International Christian University, 
3-10-2 Osawa, Mitaka-shi, Tokyo 181-8585, Japan
}
\date{\today}

\begin{abstract}
We investigate self-similar solutions 
which are asymptotic to the Friedmann universe at spatial infinity 
and contain a scalar field with potential.
The potential is required to be exponential by self-similarity.
It is found that there are two distinct 
one-parameter families of asymptotic solutions,
one is asymptotic to the proper Friedmann universe, while 
the other is asymptotic to the quasi-Friedmann universe, i.e., 
the Friedmann universe with anomalous solid angle.
The asymptotically proper Friedmann solution is 
possible only if the universe is accelerated or 
the potential is negative.
If the potential is positive, the density perturbation in the asymptotically proper Friedmann 
solution rapidly falls off at spatial infinity, while the mass perturbation is compensated.
In the asymptotically quasi-Friedmann solution, the density perturbation falls off 
only in proportion to the inverse square of the areal radius and the relative mass 
perturbation approaches a nonzero constant at spatial infinity.
The present result shows that a necessary condition holds in 
order that a self-gravitating body grows self-similarly due to 
the constant accretion of quintessence in an accelerating universe.
\end{abstract}

\maketitle


\section{Introduction}

It is now widely believed that the expansion of 
our universe got accelerated in its
early phase of evolution, which is called {\it inflation}.
Under reasonable assumptions, this inflation implies that some form 
of matter fields with largely negative pressure
may have dominated the energy of the universe.
The simplest inflation models postulate that a scalar field
with flat potential would have induced this early-phase acceleration.
On the other hand, the independent observations of 
supernovae, the cosmic microwave background, and 
the large scale structure has recently 
revealed that our universe is currently in a phase of accelerated 
expansion~\cite{astier_etal_2006,spergel_etal_2007}.
This implies that the energy of our universe is currently 
dominated by some form 
of matter fields with largely negative pressure. 
Such matter fields are termed as {\it dark energy}.
Although we do not know at present what the dark energy is, 
there are many possible candidates proposed. The first and 
simplest model is a
cosmological constant. 
Phenomenologically, the perfect fluid model with an equation of state 
$p=w\rho$ is often adopted from a data-analysis point of view, where 
$w$ might be constant or time-dependent.
On the other hand, the simplest model for varying dark energy
from a physical point of view is again a scalar field with flat potential or
possibly some other dynamical fields with 
appropriate potential~\cite{rp1988,cds1998}.
We here call such scalar field models for dark energy {\it quintessence}.
There are many variants of these varying dark energy models.

If we restrict ourselves to
the evolution of the homogeneous and isotropic universe,
the perfect fluid and the quintessence models of dark energy
play basically the same role with equivalent model functions, which is
the equation of state in the former and 
the potential in the latter.
However, once we turn our attention to inhomogeneities and/or
anisotropy, these two 
classes of models may show significant differences. 
Moreover, the response to inhomogeneous 
perturbations may distinguish the models which are degenerate in 
the homogeneous and isotropic evolution. 
Hence, it is indispensable to study inhomogeneities to distinguish the dark matter models.
Our main interest in this paper is in the interaction between 
dark energy and black holes. 

The problem of mass accretion onto black holes 
in an expanding universe was raised by Zel'dovich and
Novikov~\cite{zn1967}, where they argued that the black-hole 
mass could increase self-similarly in proportion to the cosmological time.
Although their argument was based on Newtonian gravity,
self-similar solutions also arise in general relativity 
due to the scale-free nature of the Einstein field equation.
Self-similar solutions are essentially characterized by 
functions of $z\equiv r/t$ and can describe inhomogeneous dynamics. 
They are also physically relevant because they
may describe the asymptotic behavior of 
more general solutions.
This is called {\it self-similarity hypothesis}~\cite{carr1993} and, 
in fact, this was shown to be the case in some spherically symmetric 
gravitational collapse~\cite{hm2001}.
See~\cite{cc1999,cc2005} for a recent review of self-similar solutions
and self-similarity hypothesis.
See also~\cite{barenblatt2003} for a review of self-similar solutions
in a more general context.

As we have a static black-hole solution in the Minkowski
background, which is static, it would be natural to expect that 
we may have a self-similar black hole in the power-law flat Friedmann 
background, which is self-similar.
In the study of the growth of primordial black holes~\cite{hawking1971}, 
Carr and Hawking~\cite{ch1974}
and subsequent authors~\cite{carr1976,bh1978a,bh1978b,hmc2006} 
found that, if we consider a perfect fluid with the equation of state 
$p=(\gamma-1)\rho$ for $1\le \gamma\le 2 $, there are no self-similar 
solutions which have a black-hole event horizon and are
asymptotic to the {\it proper} Friedmann solution at large distance,
but there are self-similar solutions with a black-hole event horizon 
which are only asymptotic to the Friedmann universe with some 
remaining anomaly.
It has been realized that all the latter solutions 
are only asymptotic to the Friedmann solution
with anomaly in solid angle, which are termed as
{\it asymptotically quasi-Friedmann solutions}~\cite{mkm2002,hmc2008}.

This historical problem has been recently revived by the discovery of 
the currently accelerated expansion of our universe. 
The accretion of dark energy or phantom energy onto a Schwarzschild 
black hole~\cite{bde2004,bde2005} and a 
Schwarzschild-de Sitter black hole~\cite{martin-maruno_etal_2008}
has been discussed.
The cosmological evolution partially taken into account,
it was suggested that black holes may grow self-similarly 
due to the accretion of a scalar field with potential~\cite{bm2002}.
When the cosmological evolution is fully taken into account, however,
it was shown~\cite{hmc2006} that there is {\it no} self-similar 
black-hole solution which is asymptotic to 
the {\it decelerated} Friedmann universe for a massless scalar field
and a scalar field with positive potential.
On the other hand, it has been recently found~\cite{mhc2008} that
there {\it is} a one-parameter family of self-similar 
solutions which have a black-hole event horizon 
and are asymptotic to the proper {\it accelerated} Friedmann universe
for a perfect fluid with $p=(\gamma-1)\rho$ ($0<\gamma<2/3$).
This strongly suggests that black holes can significantly grow 
due to the constant-rate 
accretion of dark energy in an accelerating universe.
However, it should be noted that
this phenomenological perfect fluid model 
for dark energy is ill-behaved
in small-scale physics~\cite{hmc2008}. 
For a scalar field with such a flat potential
that accelerates the Friedmann universe, 
it is still an open problem whether there is a self-similar 
black-hole solution which is asymptotic to the Friedmann universe.
To answer this question, it is necessary to understand the 
properties of asymptotically Friedmann 
self-similar solutions containing a scalar field with potential,
and this is investigated in the present paper.
In spite of the motivation for self-similar black holes
in the universe, the result obtained here generally applies to 
any objects which evolve in a self-similar manner and 
are embedded into the Friedmann universe containing a scalar field with potential. 

This paper is organized as follows. In Sec. II, we present a general 
formulation for self-similar solutions containing a scalar field 
with potential. In Sec. III, we rewrite the field equations 
for nonlinear perturbations from the Friedmann solution.
In Sec. IV, we find two independent one-parameter families 
of asymptotic solutions which are asymptotic to the 
Friedmann universe in different ways. 
In Sec. V, we present the physical properties of 
these asymptotic solutions.
In Sec. VI, we summarize the paper. 
We use the units, in which $c=1$.

\section{Self-similar solutions with a scalar field}
We consider a single scalar field $\varphi$ with potential $V(\varphi)$ as a matter field, 
whose stress-energy tensor is given by 
\begin{eqnarray}
T_{ab}=\varphi_{,a}\varphi_{,b}-g_{ab}\left(\frac{1}{2}\varphi_{,c}\varphi^{,c}+ V(\varphi) \right).\label{ten2}
\end{eqnarray}
As we will see later, this can accelerate the expansion of the 
Friedmann universe.
We adopt general relativity as a theory of gravity. The Einstein
equation for this system is given by
\begin{eqnarray}
R_{ab}-\frac{1}{2}g_{ab}R=
\kappa ^2\left[\varphi_{,a}\varphi_{,b}-g_{ab}
\left(\frac{1}{2}\varphi_{,c}\varphi^{,c}+V(\varphi)\right)\right],
\end{eqnarray}
where $\kappa \equiv \sqrt{8\pi G}$ and
the comma denotes the partial derivative.
The equation of motion for the scalar field is given by 
\begin{eqnarray}
\Box \varphi=\frac{d V(\varphi)}{d \varphi},
\end{eqnarray}
where $\Box$ denotes the d'Alembertian 
associated with $g_{ab}$.
We consider a spherically symmetric spacetime, in which the line element
is given by
\begin{eqnarray}
ds^2=-e^{2\Phi(t,r)}dt^2+e^{2\Psi(t,r)}dr^2+R ^2(t,r)d\Omega^2, 
\label{metric}
\end{eqnarray}
where $d\Omega^2=d\theta ^2 + \sin ^2 \theta d\phi^2$ is 
the line element on the unit sphere 
and the domain of $\theta $ and $\phi$ are $0\le \theta \le \pi$
and $0\le \phi < 2\pi$.

We assume that the spacetime is self-similar, which is defined by the existence of a vector
field $\xi^{a}$ such that
\begin{eqnarray}
\mathcal{L}_\xi g_{ab}=2g_{ab},
\end{eqnarray}
where $\mathcal{L}_\xi $ denotes the Lie derivative along $\xi^{a}$.
This vector field $\xi^{a}$ is called a homothetic Killing vector.
If $\xi^{a}$ is tilted to $(\partial/\partial t)^{a}$, 
nondimensional metric functions depend only on 
$z\equiv r/t$~\cite{ct1971} , i.e.,  
\begin{equation}
\Phi=\Phi(z), \quad \Psi=\Psi(z), \quad R =rS(z).
\end{equation}
Then, the scalar field $\varphi$ and its potential $V(\varphi)$
are of the following form~\cite{we1997}:
\begin{eqnarray}
\varphi&=&\frac{2}{\kappa \lambda }\ln r+f(z),\\
V(\varphi)&=&V_0 e^{-\kappa \lambda \varphi},
\end{eqnarray}
where $V_{0}$ and $\lambda$ are constants.
It should be noted that for simplicity we have assumed
that $r$ and $t$ are positive. 
In fact, we can recover the results for the general case
simply by replacing $r$, $t$ and $z$ with $|r|$, $|t|$ and $|z|$,
respectively.

Since we are interested in self-similar solutions perturbed from
the Friedmann universe at large distances, 
we can assume that the gradient of the 
scalar field is timelike near spacelike infinity. 
In such a case, we can take time slicing
so that $\varphi$ depends only on the time coordinate $t$, which 
we call the constant scalar field slicing.
In fact, this coordinate system is equivalent to the comoving
coordinates where only the diagonal components of the 
stress-energy tensor are nonvanishing.
In the following we choose this slicing, so that we have
\begin{equation}
 f(z)=-\frac{2}{\kappa \lambda }\ln z +\varphi_{0}
\end{equation}
and
\begin{equation}
 \varphi=\frac{2}{\kappa \lambda }\ln t+\varphi_{0},
\label{constant_scalar_field_slicing}
\end{equation}
where $\varphi_{0}$ is a constant.
If the scalar field is massless, i.e., $V_{0}=0$, 
we can simply delete $\varphi_{0}$
because only the gradient of $\varphi$ appears in the action.
If the scalar field has a potential, 
we can renormalize the constant $\varphi_{0}$ in
Eq.~(\ref{constant_scalar_field_slicing}) into the factor $V_{0}$ 
in the scalar field potential by replacing $V_{0}$ with 
$\tilde{V}_{0}$ such that
\begin{equation}
V_{0}e^{-\kappa \lambda \varphi_{0}}=\tilde{V}_{0}.
\end{equation}
Therefore, we set $\varphi_{0}=0$ in the following.

In this coordinate system, $tt$, $tr$, $rr$, and $\theta\theta$ components
of the Einstein equation, respectively, yield
\begin{eqnarray}
&& \left\{
2\frac{S^{\prime\prime}}{S}
+2\frac{S^{\prime}}{S}
+\left(1+\frac{S^{\prime}}{S}\right)^2
-2\Psi^{\prime}\left(1+\frac{S^{\prime}}{S}\right)
\right\}
-V_{z}^2\left\{2\Psi^{\prime}\frac{S^{\prime}}{S}
+\left(\frac{S^{\prime}}{S}\right)^2
\right\}
-\frac{e^{2\Psi}}{S^2}
=-\kappa ^2\left[\frac{2}{\lambda^2 \kappa ^2}V_{z}^2+z^2e^{2\Psi}V_0\right], \label{kis1}\\
&& \Phi^{\prime}\frac{S^{\prime}}{S}+\Psi^{\prime}
\left(1+\frac{S^{\prime}}{S}\right)-\frac{S^{\prime\prime}}{S}
-\frac{S^{\prime}}{S}=0,
\label{kis2} \\
&& V_{z}^2\left\{
2\frac{S^{\prime\prime}}{S}+2\frac{S^{\prime}}{S}+\left(\frac{S^{\prime}}{S}\right)^2-2\Phi^{\prime}\frac{S^{\prime}}{S}
\right\}-
\left\{
2\Phi^{\prime}\left(1+\frac{S^{\prime}}{S}\right)+\left(1+\frac{S^{\prime}}{S}\right)^2
\right\}
+\frac{e^{2\Psi}}{S^2}
=-\kappa ^2\left[\frac{2}{\lambda^2 \kappa ^2}V_{z}^2-z^2e^{2\Psi}V_0\right],
\label{kis3}\\
&& V_{z}^2\left\{
\frac{S^{\prime\prime}}{S}+\frac{S^{\prime}}{S}+\Psi^{\prime\prime}+\Psi^{\prime}+\Psi^{\prime 2}
+\left(\Psi^{\prime}-\Phi^{\prime}\right)\frac{S^{\prime}}{S}-\Phi^{\prime}\Psi^{\prime}
\right\}
-\left\{
\frac{S^{\prime\prime}}{S}+\frac{S^{\prime}}{S}+\Phi^{\prime\prime}-\Phi^{\prime}+{\Phi^{\prime}}^2+
\left(\Phi^{\prime}-\Psi^{\prime}\right)\left(1+\frac{S^{\prime}}{S}\right)-\Phi^{\prime}\Psi^{\prime}
\right\} \nonumber \\
&& =-\kappa ^2\left[\frac{2}{\lambda^2 \kappa ^2}V_{z}^2-z^2e^{2\Psi}V_0\right],
\label{kis4}
\end{eqnarray}
where the prime denotes the ordinary derivative with respect to $\ln z$ and
\begin{equation}
V_{z}\equiv z e^{\Psi - \Phi}
\label{eq:Vz_def}
\end{equation}
is the relative velocity between the constant $z$ surface 
to the constant $r$ surface. The equation of motion for the scalar field becomes
\begin{eqnarray}
-\frac{2}{\kappa \lambda }V_{z}^2
\left(\Phi^\prime - \Psi^\prime - 2\frac{S^\prime}{S}-1\right)+\kappa \lambda V_0 z^2 e^{2\Psi}=0.
\label{kis5}
\end{eqnarray}
Four of the five equations (\ref{kis1})--(\ref{kis4}) and (\ref{kis5}) 
are independent.

We derive the following two relations for later use. Adding
Eq.~(\ref{kis1}) to Eq.~(\ref{kis3}) and using Eq.~(\ref{kis2}),
we get
\begin{eqnarray}
V_{z}^2\Psi^\prime -\Phi^\prime=-\frac{2}{\lambda^2}V_{z}^2.
\label{kis+}
\end{eqnarray}
Subtracting Eq.~(\ref{kis1}) from Eq.~(\ref{kis3}) and using
Eq.~(\ref{kis2}), we get
\begin{eqnarray}
V_{z}^2\left\{\Psi^\prime \left(1+2\frac{S^\prime}{S}\right)+\left(\frac{S^\prime}{S}\right)^2 \right\}
-\left\{\Phi^\prime \left(1+2\frac{S^\prime}{S}\right)+\left(1+\frac{S^\prime}{S}\right)^2 \right\}
+\frac{e^{2\Psi}}{S^2}=\kappa ^2V_0z^2e^{2\Psi}.
\label{kis-}
\end{eqnarray}
\section{Nonlinear perturbation from the Friedmann solution}
\subsection{The flat Friedmann solution in self-similar coordinates}
The flat Friedmann spacetime is given by the following line element:
\begin{equation}
\label{533}ds^2=-dt^2+a(t)^{2}(d\bar{r}^{2}+\bar{r}^{2}d\Omega^2).
\end{equation}
The scale factor $a(t)$ satisfies the Friedmann equation
\begin{equation}
\left(\frac{\dot{a}}{a}\right)^{2}=\frac{1}{3}
\kappa ^{2}\left(\frac{1}{2}\dot{\varphi}^{2}+V(\varphi)\right),
\end{equation}
where the dot denotes the derivative with respect to $t$.
$\varphi=\varphi(t)$ satisfies the equation of motion
\begin{equation}
\ddot{\varphi}+3H\dot{\varphi}+\frac{dV}{d\varphi}=0.
\end{equation}
For these equations, we have a power-law solution
\begin{eqnarray}
a&=&a_{0}t^{\alpha },\\
\varphi&=&\frac{2}{\kappa \lambda }\ln t,
\label{friedmann_scalar_field}
\end{eqnarray}
where $a_{0}$ is a constant and $\alpha$ and $V_{0}$ are given by 
\begin{eqnarray}
\alpha &=&\frac{2}{\lambda^{2}},\\
V_{0}&=&\frac{2(6-\lambda^2)}{\kappa ^{2}\lambda^4}.
\end{eqnarray}
This is obviously compatible with Eq.~(\ref{constant_scalar_field_slicing}).
A massless scalar field formally corresponds to 
$\lambda^{2}=6$,  where $\alpha =1/3$ and 
the cosmic expansion is decelerated.
For a nontrivial potential, 
if $0<\lambda^{2}<2$, 
then the cosmic expansion is accelerated, while,  
if $\lambda^{2}>2$, 
then the cosmic expansion is decelerated. 
For $\lambda^{2}>6$, the potential becomes negative.

If $\alpha \ne 1$, relating $r$ to $\bar{r}$ through
\begin{equation}
 a_{0}\bar{r}=\frac{r^{1-\alpha }}{|1-\alpha |},
\label{rbar_r}
\end{equation}
we can rewrite the flat Friedmann solution in the standard form 
for self-similar spacetimes, where
\begin{equation}
e^{\Phi}=1, \quad e^{\Psi}=z^{-\alpha }, \quad 
S=\frac{1}{|1-\alpha |}z^{-\alpha }.\label{FRWkai}
\end{equation}
So, the power-law flat Friedmann solution is self-similar.
If $\alpha =1$, the flat Friedmann solution is still self-similar 
but the homothetic Killing vector is parallel to 
$(\partial/\partial t)^{a}$ (see e.g.~\cite{mh2005}). 
This case needs a special treatment and we 
do not consider this case in the present paper.
It should be emphasized that from Eq.~(\ref{rbar_r})
$\bar{r}\to 0$ and $\bar{r}\to \infty$ correspond to 
$r\to 0$ and $r\to \infty$, respectively, for $0<\alpha <1$,
while this is reversed for $\alpha >1$.
When we study spatial infinity in general case,
we should take the limit $z^{1-\alpha }\to \infty$
for fixed $t$ \cite{footnote}.

\subsection{Field equations for nonlinear perturbation}
Since we are interested in self-similar solutions which are 
asymptotic to the Friedmann solution, we write general 
spherically symmetric self-similar solutions in the following form:
\begin{eqnarray}
e^\Phi =e^{A(z)} , \quad e^\Psi =z^{-\alpha}e^{B(z)}, \quad 
 S=\frac{1}{|1-\alpha|}z^{-\alpha} e^{C(z)}, \quad 
\varphi=\frac{2}{\kappa \lambda }\ln t +D(z). 
\end{eqnarray}

As for the gradient of the scalar field, we get
\begin{equation}
\varphi_{,a}\varphi^{,a}=\frac{1}{t^{2}}\left[-e^{-2A}\left(\frac{2}
{\kappa\lambda}-D'\right)^{2}+z^{2\alpha-2}e^{-2B}D'^{2}\right].
\end{equation}
The above equation implies that, if $A$ and $B$ are finite and $D'$ is sufficiently small, 
we can choose the constant scalar field slicing,
where $D$ is a constant $D_{0}$. 
This is the case where the solution is 
asymptotic to the flat Friedmann solution.
Hereafter we take the constant scalar field slicing.  
We can set $D_{0}=0$.
Then we have
\begin{eqnarray}
e^\Phi =e^{A(z)} , \quad e^\Psi =z^{-\alpha}e^{B(z)}, \quad  
S=\frac{1}{|1-\alpha|}z^{-\alpha} e^{C(z)}, \quad 
\varphi=\frac{2}{\kappa \lambda }\ln t. 
\label{123}
\end{eqnarray}

Substituting the above, 
we derive the following set of ordinary differential 
equations for $A$, $B$ and $C'$:
\begin{eqnarray}
-\alpha A^\prime  + A^\prime C^\prime + (1-\alpha )B^\prime +B^\prime C^\prime-C^{\prime\prime}-(1-\alpha )C^\prime-C^{\prime 2}=0 
\label{kis22}
\end{eqnarray}
from Eq.~(\ref{kis2}), 
\begin{eqnarray}
A^\prime - B^\prime -2C^\prime -\left(3\alpha -1\right)(e^{2A}-1)=0
\label{kis55}
\end{eqnarray}
from Eq.~(\ref{kis5}), 
\begin{eqnarray}
V_{z}^{2}B^\prime  -A^\prime=0
\label{kis++}
\end{eqnarray}
from Eq.~(\ref{kis+}), and the constraint equation
\begin{equation}
(V_{z}^{2}-1)(C'^{2}-2\alpha C'-\alpha A')-2C'+(\alpha-1)^{2}
(e^{2B-2C}-1)=0
\label{eq:constraint}
\end{equation}
from Eqs.~(\ref{kis-}), (\ref{kis55}) and (\ref{kis++}), 
where $V_{z}^{2}$ is given by
\begin{eqnarray}
V_{z}^{2}= z^{2-2\alpha}e^{2B-2A}
\end{eqnarray}
from Eq.~(\ref{eq:Vz_def}).

\section{Asymptotically Friedmann solutions}
Although we are most interested in asymptotically proper 
Friedmann solutions, we
also study more general solutions to which the asymptotic scheme 
applies. So we only require that all $A$, $B$ and $C$ have finite limit
values, i.e.,
\begin{eqnarray}
A\to A_0,\;B\to B_0, \; C\to C_0
\end{eqnarray}
at spatial infinity, i.e., as $z^{1-\alpha}\to \infty$. 
Then, $A'$, $B'$, and $C'$ tend to vanish from l'Hospital's rule.
Hereafter, we always choose $A_{0}=B_{0}=0$ by
rescaling the coordinates $t$ and $r$, whereas 
$C_{0}$ may not vanish.

It is not so trivial how such asymptotic solutions are 
expanded around $z^{1-\alpha}\to\infty$. First, we note that 
only $V_{z}^{-2}$ is explicitly of higher order
in Eqs.~(\ref{kis22})--(\ref{kis++}) in the present limit.
Equation~(\ref{kis++}) then implies that $B'$ is always of higher order 
than $A'$.
If we linearize Eqs.~(\ref{kis22}) and (\ref{kis55}), we get
\begin{equation}
-\alpha A' -C''-(1-\alpha)C'=0,
\end{equation}
and
\begin{equation}
A'-2C'-2(3\alpha-1)A=0,
\end{equation}
respectively.
Then, eliminating $C'$ and $C''$, we get
\begin{equation}
A''-(5\alpha-3)A'+2(\alpha-1)(3\alpha-1)A=0.
\label{eq:A_linear}
\end{equation}
A general solution of the above equation is given by the 
linear combination of the following two independent solutions:
\begin{equation}
A=z^{3\alpha-1}
\label{eq:A_3p-1}
\end{equation}
and 
\begin{equation}
A=z^{2\alpha -2}.
\label{eq:A_2p-2}
\end{equation}
The second solution is always valid in the limit $z^{1-\alpha }\to \infty$,
while the first is valid only for $(3\alpha -1)/(1-\alpha ) <0$. 
Since $\alpha $ is positive, the first solution is
valid only for $0<\alpha <1/3$ or $\alpha >1$. 

\subsection{Asymptotically proper Friedmann solutions}
To get the full form of the first solution (\ref{eq:A_3p-1}), 
we use Eqs.~(\ref{kis22})--(\ref{kis++}) and the result is the following:
\begin{eqnarray}
A&=&A_1 z^{3\alpha -1}, \\
B&=&B_1 z^{5\alpha -3}, \\
C&=&C_0+C_1 z^{3\alpha -1}
\end{eqnarray}
in linear order, where 
\begin{eqnarray}
B_1&=&\frac{3\alpha -1}{5\alpha -3}A_1, 
\label{eq:B_1_proper}\\
C_1&=&-\frac{1}{2}A_1.
\label{eq:C_1_proper}
\end{eqnarray}

It turns out that 
we need higher order terms to see whether we can have 
nontrivial solutions and whether $C_{0}$ vanishes or not
in Eq.~(\ref{eq:constraint}).
It is cumbersome but straightforward to get higher order terms 
from Eq.~(\ref{kis22})--(\ref{kis++}). The result is
\begin{eqnarray}
A&=&A_1 z^{3\alpha -1}+A_{2} z^{5\alpha-3}
+A_{3}z^{6\alpha-2}+\cdots,  
\label{eq:A_proper}\\
B&=&B_1 z^{5\alpha -3}, 
\label{eq:B_proper}\\
C&=&C_0+C_1 z^{3\alpha -1}+
C_{2}z^{5\alpha-3}+C_{3}z^{6\alpha-2}+\cdots,
\label{eq:C_proper}
\end{eqnarray}
where the coefficients are all parametrized by $A_{1}$ as follows:
\begin{eqnarray}
A_{2}&=& \frac{\alpha}{\alpha-1}A_{1}, 
\label{eq:A_2_proper}\\
C_{2}&=& \frac{4\alpha^{2}-3\alpha+1}{2(5\alpha-3)(1-\alpha)}A_{1}, 
\label{eq:C_2_proper}\\
A_{3}&=& \frac{11\alpha-1}{8\alpha}A_{1}^{2}, 
\label{eq:A_3_proper}\\
C_{3}&=& -\frac{1}{2}A_{1}^{2}.
\label{eq:C_3_proper}
\end{eqnarray}

For $\alpha>1$, 
we consider the limit $z\to 0$. In this case, the lowest order
of Eq.~(\ref{eq:constraint}), which is of order $z^{0}$, 
just yields $C_{0}=0$.
Therefore, we get
self-similar solutions which are asymptotic to 
the proper Friedmann solution with vanishing $C_{0}$.
These solutions are termed as asymptotically proper
Friedmann solutions. 

For $0<\alpha<1/3$, the situation is more complicated.
In this case, we consider the limit $z\to \infty$.
Then, the terms of order $z^{5\alpha-3}$
in Eqs.~(\ref{eq:A_proper}) and (\ref{eq:C_proper})
get higher than those of order $z^{6\alpha-2}$.
Substituting Eqs.~(\ref{eq:A_proper})--(\ref{eq:C_proper}) 
into Eq.~(\ref{eq:constraint}), and using 
Eqs.~(\ref{eq:A_3_proper}) and (\ref{eq:C_3_proper}),
we can see that the terms of orders 
$z^{1+\alpha}$ and $z^{4\alpha}$ all cancel out.
Also in this case, 
the nontrivial lowest order, which is 
of order $z^{0}$, yields $ C_{0}=0$.
So, these self-similar solutions are asymptotically proper
Friedmann solutions.
 
For $\alpha=1/3$, 
from the linear order analysis, we find that $A=A_{1}$ and $C=C_{1}$ 
are constants, while $B$ vanishes from Eq.~(\ref{eq:B_1_proper}).
It should be noted that we may have higher order terms.
As we have set $A_{0}=0$, we can set $A_{1}=0$
by rescaling the time coordinate.  
Then, if we have higher order terms, they must 
satisfy Eq.~(\ref{eq:A_linear}) and 
this again yields $A=\mbox{const}$ and $A=z^{2\alpha-2}$.
The latter case must be included into the next case.
Hence, we can concentrate on the solution where 
$A=0$, $B=0$ and $C=C_{1}$.
In this case, we can show $C_{1}=0$ from Eq.~(\ref{eq:constraint}).
Therefore, the solution coincides with the exact Friedmann solution.

In summary, there is a one-parameter family of asymptotically 
proper Friedmann self-similar solutions for $0<\alpha<1/3$
or $1<\alpha$. There is no nontrivial asymptotically proper 
Friedmann self-similar solution for $1/3\le \alpha <1$.

\subsection{Asymptotically quasi-Friedmann solutions}
Up to the nontrivial lowest order, 
the second solution~(\ref{eq:A_2p-2}) is given by
\begin{eqnarray}
\label{454}A&=&A_1 z^{2\alpha -2}, \nonumber\\
\label{455}B&=&B_1 z^{4\alpha -4}, \nonumber\\
\label{456}C&=&C_0+C_1 z^{2\alpha -2},
\end{eqnarray}
where 
\begin{eqnarray}
B_1&=&A_1, \label{b1} \nonumber\\
C_1&=&\frac{\alpha }{1-\alpha }A_1.
\label{c1}
\end{eqnarray}
From the lowest order of Eq.~(\ref{eq:constraint}), 
which is of order $z^{0}$, we get
\begin{equation}
A_1=\frac{(\alpha -1)^2}{2\alpha (\alpha +1)}(1-e^{-2C_0}).
\label{eq:A1_C0}
\end{equation}
Higher order terms are expanded in terms of integer powers of $z^{2-2\alpha}$
and the coefficients are written by integer power of $A_{1}$.
Hence, if $C_{0}=0$, we have
$A_{1}=B_{1}=C_{1}=0$ and the solution becomes
trivial. Only if $C_{0}\ne 0$, we have a nontrivial solution.
We term these nontrivial solutions as asymptotically quasi-Friedmann
solutions. 
There is a one-parameter family of such solutions for $0<\alpha<1$
or $1<\alpha$.

\section{Physical properties of the solutions}

We have shown that there are two types of asymptotically Friedmann 
solutions with trivial and nontrivial 
asymptotic values for $C$. The first is
asymptotically proper Friedmann 
and the second is asymptotically quasi-Friedmann.
In this section we see their physical properties.

\subsection{Solid angle anomaly}

We have the following asymptotic form of the metric near spatial infinity:
\begin{eqnarray}
ds^{2}&=&-dt^{2}+z^{-2\alpha }dr^{2}+\frac{z^{-2\alpha }}{(1-\alpha )^{2}}e^{2C_{0}}r^{2}d\Omega^{2} \\
&=&-dt^{2}+a_{0}^{2}t^{2\alpha }(d\bar{r}^{2}+e^{2C_{0}}\bar{r}^{2}
d\Omega^{2}),
\end{eqnarray}
where $C_{0}=0$ and $C_{0}\ne 0$ hold for asymptotically proper Friedmann and quasi-Friedmann solutions, respectively.
If we consider a two-sphere on which $t=\mbox{const}$ and
$r=\mbox{const}$,
its area is given by 
$4\pi a_{0}^{2}t^{2\alpha } \bar{r}^{2}e^{2C_{0}}$, 
while the proper length of the radius on the 
constant $t$ hypersurface is equal to 
$a_{0}t^{\alpha }\bar{r}$. So the ratio of the 
area to the squared radius is not $4\pi$ but $4\pi e^{2C_{0}}$. 
Therefore, there is a surplus in the solid angle for $C_{0}>0$ and 
a deficit for $C_{0}<0$. 
Only for $C_{0}=0$, we have no anomaly in the solid angle. 

We can see this metric in another way.
When we consider the $\theta =\pi/2$ section, we get the line element
\begin{eqnarray}
ds^{2}&=&-dt^{2}+a_{0}^{2}t^{2\alpha }
(d\bar{r}^{2}+e^{2C_{0}}\bar{r}^{2}d\phi^{2}) \\
&=&-dt^{2}+a_{0}^{2}t^{2\alpha }
(d\bar{r}^{2}+\bar{r}^{2}d\bar{\phi}^{2}),
\end{eqnarray}
where $\bar{\phi}=e^{C_{0}}\phi$ and hence
\begin{eqnarray}
0\le \bar{\phi} < 2\pi e^{C_0}.
\end{eqnarray}
Although the above line element is the same as that 
for the $\theta=\pi/2$ section of the Friedmann
solution, the domain of the azimuthal angle is anomalous. 
In fact, there is a surplus in the azimuthal angle for $C_{0}>0$
and a deficit for $C_{0}<0$.

This kind of anomaly in the solid angle is already 
discussed in the context of static global monopoles and termed
as solid angle deficit for $C_{0}<0$~\cite{vs1994}. 
Hence, we can say that the asymptotically quasi-Friedmann solutions
are with solid angle surplus or deficit, while the asymptotically
proper Friedmann solutions are not. 
It should be noted that despite the apparent similarity with
conical singularities in cylindrically symmetric spacetimes,
the spacetime with the solid angle anomaly is not flat even locally.

\subsection{Density perturbation}
It is also interesting to get insight into the difference of
the two classes of asymptotic solutions in the
density field at spatial infinity on the constant $t$ hypersurface.
The energy density $\rho$ observed by a comoving observer 
is given by
\begin{eqnarray}
\rho\equiv n_{a}n_{b}T^{ab}
=\frac{1}{t^2}\left[\frac{1}{2}e^{-2A}
\left(\frac{2}{\kappa \lambda }\right)^2
+\frac{2(6-\lambda^2)}{\lambda^4 \kappa ^2}\right]
\simeq \frac{12}{\kappa^{2} \lambda^{4}  t^{2}}
\left[1-\frac{\lambda^{2}}{3}A\right],
\end{eqnarray}
where $n^{a}$ is a unit vector normal to the constant 
scalar field hypersurface.
Hence, the background Friedmann density $\rho_{\rm b}$,
the density perturbation $\delta\rho$, 
and the density contrast $\Delta_{\rho}$ are, respectively, given by 
\begin{eqnarray}
\rho_{\rm b}&=& \frac{12}{\kappa^{2} \lambda^{4}  t^{2}}, \\
\delta\rho &\equiv & \rho(t,r)-\rho_{\rm b}(t,r) \approx  
-\frac{4}{\kappa^{2} \lambda^{2}  t^{2}}A, \\
\Delta_{\rho} &\equiv& \frac{\delta\rho}{\rho_{\rm b}} \approx -\frac{\lambda^2}{3}A,
\end{eqnarray}
where the suffix b denotes quantities for the background Friedmann
solution
and the weak equality ``$\approx$'' denotes that the ratio of both sides 
approaches unity in the relevant limit.
The asymptotic form of the physical areal radius $R$ is given by 
\begin{eqnarray}
R =rS\approx \frac{1}{|1-\alpha |}rz^{-\alpha }e^{C_{0}}
\end{eqnarray}
for both cases. So, the fall-off of the density perturbation
in terms of the physical areal radius is given by
\begin{equation}
\delta\rho \propto -A_{1} t^{-2}z^{3\alpha -1}\propto 
-A_{1} \left(\frac{R_{\rm b}}{t}\right) ^{-\frac{3\alpha -1}{\alpha -1}}t^{-2}
\end{equation}
and
\begin{equation}
\delta\rho \propto -A_{1}t^{-2}z^{2\alpha -2}\propto -A_{1} R_{\rm b}^{-2}
\end{equation}
for asymptotically proper Friedmann and quasi-Friedmann 
solutions, respectively.
Therefore, the density perturbation rapidly falls off for 
asymptotically proper Friedmann solutions for the accelerated 
case $\alpha>1$.
It falls off as $R_{\rm b}^{-2}$ for asymptotically 
quasi-Friedmann solutions.
For asymptotically proper Friedmann solutions with $0<\alpha<1/3$,
where the potential is negative, 
the fall-off is as slow as $R_{\rm b}^{-(1-3\alpha)/(1-\alpha)}$
on the constant $t$ slice, which is much slower than for
asymptotically quasi-Friedmann solutions.

\subsection{Mass perturbation}
In spherically symmetric spacetimes, the Misner-Sharp mass $m$ is 
known to be a well-behaved quasilocal mass defined 
as~\cite{ms1964,hayward1996}
\begin{equation}
\frac{2m}{R}\equiv 1+e^{-2\Phi}R_{,t}^{2}-e^{-2\Psi}R_{,r}^{2}.
\label{eq:MS_def}
\end{equation} 
In the present formulation, this can be rewritten as
\begin{equation}
\frac{M}{S}=1+\frac{1}{(1-\alpha)^{2}}z^{2-2\alpha}e^{2C-2A}
(-\alpha+C')^{2}
-\left(1+\frac{C'}{1-\alpha}\right)^{2}e^{2C-2B},
\end{equation}
where $M$ is the nondimensional mass defined by
\begin{equation}
M\equiv \frac{2m}{r}.
\end{equation}

For the flat Friedmann solution, this quantity becomes
\begin{equation}
\left(\frac{2m}{R}\right)_{\rm b}=
\frac{\alpha^{2}}{(1-\alpha)^{2}}z^{2-2\alpha}=\alpha^{2}\left(\frac{R_{\rm b}}{t}\right)^{2}.
\end{equation}
The perturbation for this quantity 
\begin{equation}
\delta \left(\frac{2m}{R}\right) \equiv \left(\frac{2m}{R}\right)(z)-
\left(\frac{2m}{R}\right)_{\rm b}(z)
\end{equation}
is given by
\begin{equation}
\delta \left(\frac{2m}{R}\right)\approx -\frac{\alpha}{(1-\alpha)^{2}}A_{1}
z^{1+\alpha}=-\alpha |1-\alpha|^{(3\alpha-1)/(1-\alpha)}A_{1}\left(\frac{R_{\rm b}}{t}\right)^{(1+\alpha)/(1-\alpha)}
\end{equation}
for asymptotically proper Friedmann solutions and 
\begin{equation}
\delta \left(\frac{2m}{R}\right)\approx 
\frac{\alpha^{2}}{(1-\alpha)^{2}}(e^{2C_{0}}-1)z^{2-2\alpha}
=\alpha^{2}(e^{2C_{0}}-1)\left(\frac{R_{\rm b}}{t}\right)^{2}
\end{equation}
for asymptotically quasi-Friedmann solutions.
Hence, the ratio of the perturbation to the background value 
\begin{equation}
\Delta \equiv \delta \left(\frac{2m}{R}\right)(z)/ 
\left(\frac{2m}{R}\right)_{\rm b}(z)
\end{equation}
is given by
\begin{equation}
\Delta 
\approx  -\frac{1}{\alpha}A_{1}z^{3\alpha-1}
=-\frac{|1-\alpha|^{(3\alpha-1)/(1-\alpha)}}
{\alpha}A_{1}\left(\frac{R_{\rm b}}{t}\right)^{(3\alpha-1)/(1-\alpha)}
\end{equation}
and 
\begin{equation}
\Delta \approx  e^{2C_{0}}-1
\end{equation}
for asymptotically proper Friedmann and quasi-Friedmann solutions,
respectively.

It should be noted that since $m$ and $R$ are nonlinearly 
perturbed for asymptotically quasi-Friedmann solutions,
this ratio depends on whether we compare the perturbation at the 
same $r/t$ or at the same $R/t$.
Noting that $S$ is also perturbed, the mass perturbation 
$\delta M(z)\equiv M(z)-M_{\rm b}(z)$ is directly given by
\begin{equation}
\Delta_{M}\equiv \frac{\delta M}{M_{\rm b}}=(1+\Delta)e^{C}-1.
\end{equation} 
For asymptotically proper Friedmann solutions, 
$\Delta_{M}$ is calculated as
\begin{equation}
\Delta_{M}\approx -\frac{\alpha+2}{2\alpha}A_{1}z^{3\alpha-1}
\propto -A_{1}\left(\frac{R_{\rm b}}{t}\right)^{(3\alpha-1)/(1-\alpha)}.
\end{equation}
Hence, the relative mass perturbation $\Delta_{M}$ tends to vanish 
at spatial infinity for both $\alpha>1$ and $0<\alpha<1/3$.
This also implies that the mass perturbation $\delta m$ itself 
tends to vanish for $\alpha>1$ but diverge  for $0<\alpha<1/3$
as it is proportional to $(R_{\rm b}/t)^{2/(1-\alpha)}t$.
For asymptotically quasi-Friedmann solutions,
$\Delta_{M}$ is calculated as
\begin{equation}
\Delta_{M}\approx e^{3C_{0}}-1.
\end{equation}
Hence, the relative mass perturbation tends to be constant.
This is directly related to the solid angle anomaly.
If $\Delta_{M}$ is positive (negative), there is a solid 
angle surplus (deficit). 
This situation is apparently opposite to the case of global monopoles,
where the positive (negative) mass density implies deficit (surplus)
in the solid angle. This is due to the fact that the Misner-Sharp mass
is dominated by the first and third terms on the right-hand side 
of Eq.~(\ref{eq:MS_def})
for the static configuration, while it is by the second term, i.e.,
the kinematic term, for 
the flat Friedmann solution.
The mass perturbation $\delta m$ itself diverges as $(R/t)^{3}t$
for both $0<\alpha<1$ and $1<\alpha$.

So, in order to have an asymptotically proper Friedmann solution
from an accelerated Friedmann universe,
for which the potential is positive, we only need to perturb a finite amount of mass and
the mass perturbation is compensated at spatial infinity.
In contrast, in order to have an asymptotically quasi-Friedmann solution from the 
Friedmann solution, we need to perturb an infinite amount of mass 
and the mass perturbation remains at spatial infinity.
This is also the case so as to  
have an asymptotically proper Friedmann solution
from the decelerated Friedmann solution, for which the potential 
is negative. 
This suggests that asymptotically proper accelerated Friedmann solutions
are physically acceptable as nonlinearly perturbed
solutions from the Friedmann solution by some classical mechanism. 
This also suggests that any classical perturbation 
mechanism will not perturb a Friedmann universe to 
a quasi-Friedmann universe. Only through quantum fluctuations,
it might be possible to have a quasi-Friedmann solution
because an infinite amount of perturbed mass must extend 
in scales much larger than the Hubble horizon at any epoch.
On the other hand,  
whether the perturbed mass is compensated or remains
or even diverges at spatial infinity, the present 
perturbation scheme is still completely applicable
for these asymptotic solutions.

\subsection{Comparison with a perfect fluid with $p=(\gamma-1)\rho$}
For the Friedmann solution, a scalar field with exponential 
potential and a perfect fluid with $p=(\gamma-1)\rho$ play a completely
equivalent role. In the spatially flat case, they are related with 
the following relation:
\begin{equation}
\alpha=\frac{2}{\lambda^{2}}=\frac{2}{3\gamma}.
\end{equation}
So, the accelerated expansion is possible 
if $0<\lambda^{2}<2$ for the scalar field and 
if $0<\gamma<2/3$ for the perfect fluid.
However, once we admit perturbations from a uniform distribution,
the two systems get very different. 

For example, a scalar wave propagates at the speed of light in the 
short wave length limit in the scalar field system 
even in the presence of potential. 
In contrast, in the perfect fluid system with the equation of state $p=(\gamma-1)\rho$, a sound wave propagates at the 
sound speed $\sqrt{\gamma-1}$ for $1<\gamma\le 2$ and,
in fact, there is no sound wave but instability 
in the short wave length limit for $0<\gamma< 1$~\cite{hmc2008}.

Also in the perfect fluid system 
with $p=(\gamma-1)\rho$, there are two independent
one-parameter families of solutions which are asymptotic 
to the Friedmann solution~\cite{hmc2008,mkm2002}. 
One is asymptotically proper Friedmann solutions at spatial
infinity and the other is asymptotically quasi-Friedmann solutions
at spatial infinity. The latter is valid for both the 
accelerating ($0<\gamma<2/3$ or $\alpha>1$)
and decelerating ($2/3<\gamma<2$ or $1/3<\alpha<1$) cases, while the former is 
only valid for the accelerating case.
Hence, the situation is exactly parallel to that in the 
scalar field case. This is a very unexpected result 
because we do admit inhomogeneity when we consider asymptotic solutions.

In the perfect fluid analysis, 
the strongly decelerated case, $0<\alpha<1/3$ or $\gamma>2$,
has not been analyzed because causality is violated in 
such a model. 
In the present analysis, on the other hand, 
since the scalar field with negative potential is 
acceptable from a causal point of view,
we have included this case and found interesting features that
both asymptotically proper and quasi-Friedmann solutions exist
and that the asymptotically proper Friedmann solutions are
very different from those for the accelerated case.

For a perfect fluid with $1\le \gamma \le 2$, there is no 
self-similar solution which has a black-hole event horizon 
and is asymptotic to the proper Friedmann solution at 
spatial infinity. However, 
for a perfect fluid with $0<\gamma<2/3$, 
there is a one-parameter family of 
asymptotically proper Friedmann solutions. In fact, the numerical 
integration has revealed that there is a one-parameter family of 
self-similar solutions among them which contain a black-hole event horizon~\cite{mhc2008}. 
To implement the numerical integration in that case, it is highly advantageous
that the system of ordinary differential equations has no 
critical surface because there is no 
propagation of sound wave. 
Also in the scalar field case, one might guess that 
the existence of asymptotically proper Friedmann solutions
suggests the existence of self-similar black-hole solutions 
belonging to this class.
However, the scalar field system has a
critical surface coinciding with a similarity horizon, where $V_{z}=1$.
This makes the problem complicated because this 
could possibly increase the number of self-similar solutions drastically 
as such a critical surface may admit weak discontinuity.
In this connection, we should also note that 
because of the critical surface,
the power-law flat Friedmann solution containing a scalar field with 
potential is unstable for $4<\lambda^{2}<6$
against weak discontinuity, i.e., the 
kink mode at a particle horizon~\cite{mh2004}.
It is however stable for $0<\lambda^{2}<4$,
marginally stable for $\lambda^{2}=4$.
This kink instability might be related to the physical relevance of 
self-similar solutions.

\section{Summary}

We have considered self-similar nonlinear perturbation 
from the Friedmann solution and investigated the asymptotic properties of 
spherically symmetric self-similar solutions 
containing a scalar field with potential and approaching the 
flat Friedmann solution at spatial infinity.
The potential is restricted from self-similarity 
to be exponential with the steepness parameter $\lambda$.
This is motivated by the fact that a scalar field with sufficiently
flat ($0<\lambda^{2}<2$) potential enables the universe to expand 
with acceleration and hence acts as quintessence.

If the potential is so flat, i.e., $0<\lambda^{2}<2$ that the Friedmann
universe expands with acceleration, we have found that 
there is a one-parameter family of self-similar solutions 
which are asymptotic to the proper 
Friedmann solution at spatial infinity.
Furthermore, we have found that there is also a one-parameter 
family of self-similar solutions which are asymptotic
to the Friedmann solution but with some 
anomaly in solid angle. Such solutions are called  
asymptotically quasi-Friedmann solutions.

If the potential is steep, i.e. $\lambda^{2}>2$, 
we have the Friedmann universe decelerated. 
Even in such a potential, we have found a 
one-parameter family of self-similar solutions which 
are asymptotically quasi-Friedmann solutions. 
However, we have also shown that there is no 
nontrivial asymptotically proper 
Friedmann self-similar solution in this case
as long as the potential is positive.
We should note that it was already shown that there is no 
self-similar solution which contains a black-hole event horizon 
and is asymptotically proper Friedmann or 
quasi-Friedmann
for a scalar field with positive potential inducing 
the decelerating expansion~\cite{hmc2006}. 

Our analysis includes the case of a massless scalar field, where 
the flat Friedmann universe is decelerated.
In this case, we have found that 
there is a one-parameter family of asymptotically quasi-Friedmann
self-similar solutions, while 
there is no nontrivial asymptotically proper Friedmann self-similar solution. 
We should also note that it was already shown that there is no 
self-similar solution which contains a black-hole event horizon 
and is asymptotically proper Friedmann or quasi-Friedmann 
for a massless scalar field~\cite{hmc2006}. 

Our analysis also includes the case where the potential
is negative. In such a case, the Friedmann universe is 
strongly decelerated. 
We have found that there are both 
one-parameter families of asymptotically quasi-Friedmann
self-similar solutions and asymptotically proper Friedmann
self-similar solutions.
The latter is very different
in density and mass perturbations
from that for the positive potential.

We have shown that the perturbed mass
is finite for asymptotically proper Friedmann solutions 
as long as the potential is positive.
In contrast, it is infinite for asymptotically quasi-Friedmann 
solutions.
This suggests that asymptotically proper Friedmann solutions 
are physically more acceptable as solutions perturbed  
from the Friedmann universe through some causal mechanism 
than asymptotically quasi-Friedmann solutions.
Although asymptotically proper Friedmann solutions 
are possible even if the potential is negative,
the perturbed mass is infinite there.

Although we have found the above interesting properties of 
self-similar solutions containing a scalar field with potential,
it is still an open question
whether there is a self-similar black-hole 
solution which is asymptotically proper or quasi-Friedmann.
We need possibly a numerical analysis based on the 
present asymptotic analysis, as 
it has revealed the existence of self-similar black-hole solutions 
for a perfect fluid with the equation 
of state $p=(\gamma-1)\rho$ ($0<\gamma<2/3$)~\cite{hmc2008,mhc2008}.
Although spherically symmetric self-similar 
solutions with scalar fields have been also investigated 
in a dynamical systems approach~\cite{ch2002,ct2001,cg2000}, 
no definite answer to the existence of black-hole solutions 
has been reported yet.
It is an important future work to answer whether there is a 
self-similar black-hole solution in quintessential cosmology 
and, if it exists, to 
study the physical properties of such a black-hole solution.

\acknowledgments

The authors would like to thank B.~J. Carr and R. Tavakol
for useful comments.
TH and HM were supported by the Grant-in-Aid for Scientific
Research Fund of the Ministry of Education, Culture, Sports, Science
and Technology, Japan (Young Scientists (B) 18740144 and 18740162),
respectively.
HM was also supported by the Grant No. 1071125 from FONDECYT (Chile).
The Centro de Estudios Cient\'{\i}ficos (CECS) is funded by the Chilean 
Government through the Millennium Science Initiative and the Centers of
Excellence Base Financing Program of Conicyt. CECS is also supported by 
a group of private companies which at present includes Antofagasta 
Minerals, Arauco, Empresas CMPC, Indura, Naviera Ultragas, and 
Telef\'{o}nica del Sur.


\begin{thebibliography}{99}
\bibitem{astier_etal_2006}
P.~Astier {\it et al.},
Astron. Astrophys. {\bf 447}, 31 (2006).
\bibitem{spergel_etal_2007}
D.~N.~Spergel et al.,
Astrophys. J. Suppl. {\bf 170}, 377 (2007). 
\bibitem{rp1988}
B.~Ratra and P.~J.~E.~Peebles, Phys. Rev. D{\bf 37}, 3406 (1988).
\bibitem{cds1998}
R.~R.~Caldwell, R.~Dave and P.~J.~Steinhardt, Phys. Rev. Lett. {\bf 80},
	 1582 (1998).
\bibitem{zn1967}
   Ya.~B.~Zel'dovich and I.~D.~Novikov,
   Sov. Astron. {\bf 10}, 602 (1967).
\bibitem{carr1993}
B.~J.~Carr, preprint prepared for but omitted from {\it The Origin of 
Structure in the Universe}, ed. E. Gunzig and P. Nardone (Kluwer, 1993).
\bibitem{hm2001} T.~Harada and H.~Maeda, 
Phys. Rev. D{\bf 63}, 084022 (2001).
\bibitem{cc1999}
B.~J.~Carr and A.~A.~Coley, Class. Quant. Grav. {\bf 16}, R31 (1999).
\bibitem{cc2005}
B.~J. Carr and A.~A.~Coley, Gen. Rel. Grav. {\bf 37}, 2165 (2005).
 \bibitem{barenblatt2003}
G.~I.~Barenblatt, {\it Scaling}, (Cambridge University Press, Cambridge, 2003).
 \bibitem{hawking1971}
S.~W. Hawking, Mon. Not. R. Astron. Soc. {\bf 152}, 75 (1971).
 \bibitem{ch1974}
B.~J. Carr and S.~W. Hawking, Mon. Not. R. Astron. Soc. {\bf 168}, 399 (1974).
\bibitem{carr1976}
B.~J.~Carr,
Ph.D. thesis, Cambridge University (1976).
\bibitem{bh1978a}
   G.~V.~Bicknell and R.~N.~Henriksen,
   Astrophys. J. {\bf 219}, 1043 (1978).
\bibitem{bh1978b}
   G.~V.~Bicknell and R.~N.~Henriksen,
   Astrophys. J. {\bf 225}, 237 (1978).
\bibitem{hmc2006}
T.~Harada, H.~Maeda and B.~J.~Carr,
Phys. Rev. D{\bf 74}, 024024 (2006).
\bibitem{mkm2002}
H.~Maeda, J.~Koga and K.-i.~Maeda, Phys. Rev. D{\bf 66},
087501 (2002).
 \bibitem{hmc2008}
T.~Harada, H.~Maeda and B.~J.~Carr, Phys. Rev. D{\bf 77}, 024022 (2008).
\bibitem{bde2004}
E.~Babichev, V.~Dokuchaev and Yu.~Eroshenko,
Phys. Rev. Lett.
{\bf 93}, 021102 (2004).
\bibitem{bde2005}
E.~Babichev, V.~Dokuchaev and Yu.~Eroshenko,
J. Exp. Theor. Phys.
{\bf 100}, 528 (2005).
\bibitem{bm2002}
R.~Bean and J.~Magueijo,
Phys.~Rev.~D{\bf 66}, 063505 (2002). 
\bibitem{martin-maruno_etal_2008}
P.~Mart\'in-Moruno, A.-E.~L.~Marrakchi, S.~Robles-P\'erez and
	P.~F.~Gonz\'alez-D\'iaz, Report No. arXiv:0803.2005v1.
 \bibitem{mhc2008}
H.~Maeda, T.~Harada and B.~J.~Carr, Phys. Rev. D{\bf 77}, 024023 (2008).
\bibitem{ct1971}
M.~E.~Cahill and A.~H.~Taub, Commun. Math. Phys. {\bf 21}, 1 (1971).
\bibitem{we1997}
J.~Wainwright and G.~F.~R.~Ellis, {\it Dynamical Systems in Cosmology}, 
(Cambridge University Press, Cambridge, 1997).
\bibitem{mh2005}
H.~Maeda and T.~Harada, in {\it General Relativity Research
Trends, Horizons in World Physics Vol. 249}, edited by
R. Albert (Nova Science Publishers, New York, 2006),
p. 123; Report No. gr-qc/0405113.
\bibitem{footnote}
The fact that $\bar{r}\to \infty$ corresponds to 
$r\to 0$ for an accelerated Friedmann universe 
is overlooked in~\cite{cc1999,mkm2002} and correctly considered 
in~\cite{hmc2008,mhc2008}.
\bibitem{vs1994} 
A.~Vilenkin and E.~P.~S.~Shellard,
{\it Cosmic Strings and Other Topological Defects}, 
(Cambridge University Press, Cambridge, 1994).
\bibitem{ms1964}
C.~W.~Misner and D.~H.~Sharp,
Phys. Rev. {\bf 136}, B571 (1964).
\bibitem{hayward1996}
S.~A.~Hayward, 
Phys. Rev. {\bf D53}, 1938 (1996).
\bibitem{mh2004}
H.~Maeda and T.~Harada, Phys. Lett. B{\bf 607}, 8 (2005).
\bibitem{cg2000}
A.A.~Coley and M.~Goliath, 
Class. Quant. Grav. {\bf 17}, 2557 (2000).
\bibitem{ct2001}
A.~A.~Coley and T.D.~Taylor, 
Class. Quant. Grav. {\bf 18}, 4213 (2001).
\bibitem{ch2002}
A.~A.~Coley and Y.~He,
Class. Quant. Grav. {\bf 19}, 3901 (2002).
\end{thebibliography}
\end{document}